# Macroscopic quantum superposition of spin ensembles with ultra-long coherence times via superradiant masing

Liang Jin[1], Sen Yang[2], Jörg Wrachtrup[2] & Ren-Bao Liu[1]

1. *Department of Physics, Centre for Quantum Coherence & Institute of Theoretical Physics, The Chinese University of Hong Kong, Shatin, New Territories, Hong Kong, China*

2. *3 Physikalisches Institut, University of Stuttgart, Pfaffenwaldring 57, 70569 Stuttgart, Germany*

**Macroscopic quantum phenomena such as lasers, Bose-Einstein condensates, superfluids, and superconductors are of great importance in foundations [1-3] and applications [4-6] of quantum mechanics. In particular, quantum superposition of a large number of spins in solids is highly desirable for both quantum information processing [7-16] and ultrasensitive magnetometry [17, 18]. Spin ensembles in solids, however, have rather short collective coherence time (typically less than microseconds [9, 14]). Here we demonstrate that under realistic conditions it is possible to maintain macroscopic quantum superposition of a large spin ensemble (such as about $\sim 10^{14}$ nitrogen-vacancy center electron spins in diamond) with an extremely long coherence time $\sim 10^8$ sec under readily accessible conditions. The scheme, following the mechanism of superradiant lasers [19, 20], is based on superradiant masing due to coherent coupling between collective spin excitations (magnons) and microwave cavity photons. The coherence time of the macroscopic quantum superposition is the sum of the magnon life time and the cavity lifetime, further elongated by the total number of coherent magnons and photons, which have macroscopic values when masing occurs. The macroscopic quantum coherence of spin ensembles can be exploited for magnetometry with sensitivity $\sim 10\,\text{fT}\cdot\text{Hz}^{-1/2}$. The long-living collective states of spin ensembles in solids will provide a new platform for studying macroscopic quantum phenomena and are useful for quantum technologies.**

A most well-known macroscopic quantum phenomenon is laser, where long coherence times are achieved by collective enhancement. To illustrate such collective enhancement, let us consider first the spontaneous emission of photons from individual atoms (Fig. 1a). The photons emitted at different times have different phases, randomly shifted by an amount in the order of π. Therefore, the coherence time of the spontaneously emitted photons is the lifetime ($\tau_a$) of the emitters in their excited states. If the atoms are placed between high-reflectance mirrors (which form an optical cavity) the photons would be reflected between the mirrors for many times before leaving the cavity after time $\tau_c$ (called cavity lifetime) (Fig. 1b). The photon coherence time will be $\tau_c$ since photons separated by a time longer than $\tau_c$ should come from different emitters and hence have random relative phases. When the atoms in the cavity are inverted, the photons in the cavity will stimulate emission of the atoms, resulting in lasing (Fig. 1c). The photons generated by stimulated emission have the same phase. Now that the random phase is shared by a macroscopic number ($n_c$) of photons, the coherence time of the laser is $\sim n_c \tau_c$.

The coherence times of lasers can be further enhanced by collective excitations in the atoms. In lasing considered above, the lifetime of the excited states of atoms is assumed much shorter than the cavity lifetime ($\tau_a \ll \tau_c$), corresponding to the good cavity limit). In the case that the atoms have long lifetime $\tau_a \sim \tau_c$ or $\tau_a \gg \tau_c$, the photon can be reabsorbed by the atoms and the coherence stored there for a time $\sim \tau_a$ (similar to atomic memory), so the laser coherence time is $\sim n_c (\tau_a + \tau_c)$, enhanced by the atomic coherence [21, 22]. Furthermore, there could be collective excitations in the atoms via the so-called superradiance effect, in which all the atoms are coupled to the same photon modes and can be excited to a collective superposition mode (similar to spin waves or magnons) (Fig. 1d). In such superradiant lasing [1920], the quantum coherence can be stored in both the atomic collective mode and the cavity mode. The random phase is shared by photons and magnons, so if the number of "magnons" is $n_a$, the superradiant laser has coherence time

$$T_{\text{coh}} \sim (n_a + n_c)(\tau_a + \tau_c), \qquad (1)$$

where $\tau_a$ is the lifetime of the atomic collective mode instead of the coherence time of single atoms. Thus long coherence times of atomic collective modes can be employed to enhance the laser coherence times even in the bad cavity limit ($\tau_c \ll \tau_a$). It has been theoretically proposed [19] and experimentally verified [20] that superradiant lasing can be realized with few photons in cavities (even <<1), where the long coherence times are enabled by the large number of collective excitations ($n_a \gg n_c$), in contrast to conventional lasers.

Here, we propose to exploit superradiant lasing to overcome a grand challenge in quantum science and technology – to greatly enhance the coherence times of collective modes of spin ensembles by coupling them to a high-quality microwave cavity and driving the coupled system into superradiant masing (lasing in the microwave waveband). The coherence time is enhanced by the large number of photons coherently coupled to the magnon excitations in the spin ensembles (with $n_c \gg n_a$). Spin ensembles, bearing the feature of enhanced coupling to resonators and external fields, have been considered as a promising candidate for quantum technologies (such as quantum interfaces, quantum memories, and ultrasensitive sensors) [7-18]. Also, the collective modes of spin ensembles may provide a controllable platform for studying macroscopic quantum phenomenon. However, the lifetimes of the collective modes of spin ensembles are too short due to many-body interactions and inhomogeneous broadening [9, 14]. Here, using nitrogen-vacancy (NV) center spin ensembles in diamond [23] as a specific example, we demonstrate that through the superradiant masing an ensemble of $\sim 10^{14}$ spins can be sustained in a macroscopic quantum superposition with extremely long coherence times $\sim 10^8$ sec, in sharp contrast to the microsecond-scale lifetime of the collective spin excitation. As an example of application, such long coherence times render the sensitivity of magnetometry to be $\sim 10\,\text{fT} \cdot \text{Hz}^{-1/2}$.

We consider an ensemble of NV center spins in diamond [23] resonantly coupled to a high quality Febry-Pérot microwave cavity (Fig. 2a). Note that many other types of solid-state spin ensembles and microwave cavities [8-14, 24, 25] may be considered for implementing the proposal in this paper. The three states of an NV center spin $|0\rangle$ and $|\pm1\rangle$ have a zero-field splitting about 2.87 GHz [23] (Fig. 2b). The NV centers can be optically pumped to the state $|0\rangle$ [23]. A moderate external magnetic field can split the $|\pm1\rangle$ and shift the $|-1\rangle$ state below $|0\rangle$ so that the spins can be inverted by optical pumping. The transition frequency $\omega_S$ between the spin ground state $|g\rangle \equiv |-1\rangle$ and the exited state $|e\rangle \equiv |0\rangle$ is tuned near resonant with the cavity frequency $\omega_c$. The diamond is placed at the center of the cavity so that the cavity mode couples to the spins with the interaction Hamiltonian $H_I = \sum_{j=1}^{N} g_j \left( \hat{a}\hat{s}_j^+ + \hat{a}^\dagger \hat{s}_j^- \right)$, where $\hat{a}$ annihilates a photon, $\hat{s}_j^+ \equiv |e\rangle_{jj}\langle g|$ is the raising operator of the $j$-th spin, $\hat{s}_j^- = \left(\hat{s}_j^+\right)^\dagger$, and $g_j$ is the coupling constant. By defining the collective spin raising/lowering operators $\hat{S}_\pm \equiv g^{-1} \sum_{j=1}^{N} g_j \hat{s}_j^\pm$ and the collective coupling constant $g \equiv \sqrt{N^{-1} \sum_{j=1}^{N} g_j^2}$, the interaction Hamiltonian can be written as $H_I = g\left( \hat{a}\hat{S}_+ + \hat{a}^\dagger \hat{S}_- \right)$. Without changing the essential results, we assume the spin-photon coupling is a constant, i.e., $g_j = g$. That simplifies the collective operators as $\hat{S}_\pm \equiv \sum_{j=1}^{N} \hat{s}_j^\pm$. The collective spin operators satisfy the commutation relation $\left[\hat{S}_-, \hat{S}_+\right] = \sum_{j=1}^{N} \left( |e\rangle_{jj}\langle e| - |g\rangle_{jj}\langle g| \right) \equiv \hat{S}_z$. When masing occurs, the spin polarization (or population inversion) $S_z \equiv \langle \hat{S}_z \rangle$ is a macroscopic number [ $\sim O(N)$ ] while the fluctuation $\delta\hat{S}_z \equiv \hat{S}_z - S_z \sim O(N^{1/2})$ is much smaller. Therefore, $\hat{b}^\dagger \equiv \hat{S}_+ / \sqrt{S_z}$ can be interpreted as the creation operator of a collective magnon mode with $[\hat{b}, \hat{b}^\dagger] \cong 1$. The magnon creation operator generates coherent superposition states in the spin ensemble. For example, from a fully polarized spin state, the one magnon state is $\hat{b}^\dagger |g\rangle_1 |g\rangle_2 \cdots |g\rangle_N = \sqrt{1/N} \sum_{j=1}^{N} |g\rangle_1 \cdots |g\rangle_{j-1} |e\rangle_j |g\rangle_{j+1} \cdots |g\rangle_N$. In the masing state, the photon and the magnon modes, coherently coupled to each other, are both in coherent states, with macroscopic amplitudes. With the number of coherent magnons $n_S \equiv \langle \hat{b}^\dagger b \rangle = \langle \hat{S}_+ \hat{S}_- \rangle / S_z \sim O(N)$, the spins are in a macroscopic quantum superposition state maintained by the superradiant masing process.
4_

Now we describe the superradiant masing process. The key to masing is to invert the spin populations. This can be achieved for NV center spins by optical pumping [23], in which a laser brings the center from the states $|0\rangle$ and $|\pm 1\rangle$ to optically excited states, and then the center returns back to the spin state with higher probability to the state $|0\rangle$, providing an effective incoherent pumping from the state $|g\rangle \equiv |-1\rangle$ to $|e\rangle \equiv |0\rangle$ (see Fig. 2b). The pumping rate $w$ can be tuned by varying the pumping laser intensity, up to $\sim 10^7$ sec$^{-1}$ [26]. The cavity mode has a decay rate determined by the cavity quality factor $Q$, $\kappa_c = 2\tau_c^{-1} = \omega_c/Q$, due to leakage of photons (i.e., maser emission). The decay of the magnon mode is caused by various mechanisms. First, the spin relaxation ($T_1$ process caused by phonon scattering and resonant interaction between spins) contributes a decay rate $\gamma_{eg} = 1/T_1$. Second, the individual spins experience local field fluctuations due to interaction with nuclear spins, coupling to other NV and nitrogen center spins, and fluctuation of the zero-field splitting. Such local field fluctuation induces random phases $\varphi_j$ to individual spins, making a superposition state, e.g., a one-magnon state, to $\sqrt{1/N}\sum_{j=1}^{N} e^{i\varphi_j}|g\rangle_1 \cdots |g\rangle_{j-1}|e\rangle_j|g\rangle_{j+1} \cdots |g\rangle_N$, which has decaying overlap with the original magnon state with the phase randomness increasing. So the local field fluctuation leads to decay of the magnon mode with the rate $2/T_2^*$, where $T_2^*$ is the dephasing time of the spin ensembles. Finally the optical pumping, being incoherent, gives a decay rate $w$ of the magnon mode. The total decay rate of the magnon mode is thus $\kappa_S = w + 2/T_2^* + \gamma_{eg}$. The quantum dynamics of the coupled magnons and photons are described by the Langevin equations [22] for the magnon and photon operators $\hat{a}$ and $\hat{S}_\pm$ and the spin operators $\hat{N}_{e/g} \equiv \sum_{j=1}^{N}|e/g\rangle_{jj}\langle e/g|$ and $\hat{S}_z$ (see Methods for details).

For a specific system, we consider a diamond sample of volume $V_{NV} = 3\times 3\times 0.5$ mm$^3$ with the NV center concentration $\rho_{NV} = 10^{17}$ cm$^{-3}$, natural abundance (1.1%) of $^{13}$C nuclear spins, and nitrogen (P1) center concentration about 5 ppm [18, 27]. The ensemble spin decoherence time (which is mostly caused by the dipolar interaction with the nearby P1 center electron spins and $^{13}$C nuclear spins, and the zero-field splitting fluctuation) is $T_2^* = 0.5$ μs [16]. The number of NV centers coupled to the cavity mode is $N = 0.375 \times 10^{14}$. The external magnetic field 2100 G results in $\omega_S/2\pi \approx 3$ GHz. The microwave cavity has length $L \approx 50$ mm and has its frequency resonant with the magnon, $\omega_c = \omega_S$. The spin-photon coupling is about $g/2\pi = 0.02$ Hz for the effective cavity mode volume $V_{\text{eff}} \approx 2.5$ cm$^3$ [24]. At low temperature (< 10 K), the spin relaxation is mainly caused by resonant interaction between NV center spins and $\gamma_{\text{eg}} = 1/T_1 \approx 0.05$ sec$^{-1}$ corresponding to the NV center ~concentration. At room temperature, the phonon scattering dominates the spin relaxation and $\gamma_{\text{eg}} \approx 200$ sec$^{-1}$. The number of thermal photons inside the cavity is $n_{\text{th}} \approx 0.43$ at 120 mK.

The quantum Langevin equations can be easily solved under the masing condition. When masing occurs, the quantum operators can be approximated as their expectation values, i.e., $\hat{S}_\pm \approx S_\pm$, $\hat{a} \approx a$, $\hat{N}_{e/g} \approx N_{e/g}$, and $\hat{S}_z \approx S_z$. By dropping the small quantum fluctuations, we reduce the quantum Langevin equations to classical equations for the operator expectation values (see Methods for details). Under the exact resonance condition ($\omega_S = \omega_c$), the steady state solution is

$$S_z = \kappa_S \kappa_c / (4g^2),$$
$$S_- = i\sqrt{S_z \left( \frac{w - \gamma_{\text{eg}}}{2\kappa_S} N - \frac{w + \gamma_{\text{eg}}}{2\kappa_S} S_z \right)}, \quad (2)$$
$$a = \sqrt{\frac{w - \gamma_{\text{eg}}}{2\kappa_c} N - \frac{w + \gamma_{\text{eg}}}{2\kappa_c} S_z}.$$

Note that to have a population inversion (spin polarization) $S_z \sim O(N)$, the pump rate should scale with the total number of spins as $w \sim O(N)$. The fact that the photon number $n_c = a^* a = \frac{w - \gamma_{\text{eg}}}{2\kappa_c} N - \frac{w + \gamma_{\text{eg}}}{2\kappa_c} S_z > 0$ leads to the masing condition

$$\kappa_c < \frac{4g^2}{\kappa_S} \frac{w - \gamma_{\text{eg}}}{w + \gamma_{\text{eg}}} N. \quad (3)$$

Firstly, this condition means the pumping rate $w$ has to be greater than the spin relaxation rate $\gamma_{eg}$ to maintain population inversion. Secondly, the cavity quality $Q$ has to be above a threshold to have a sufficient number of photons to maintain the phase correlations between the spins. Stronger spin-photon coupling, longer magnon lifetime, or a larger number of spins can reduce this threshold of cavity $Q$ factor. Thirdly, the magnon decay rate $\kappa_S$ should be kept below the maximal collective emission rate of photons $4Ng^2/\kappa_c$, otherwise over repumping would fully polarize the spins, making the spin-spin correlation vanish ($S_z \to N$ and $S_- \to 0$).

Emergence of macroscopic quantum superposition is evidenced by macroscopic values of the spin polarization, the photon amplitude, and the magnon amplitude under the masing condition. We calculated the spin polarization and the photon and magnon numbers using the higher order equations of the correlation functions (see Mehtods), which apply to both masing and incoherent regimes. The calculated results of $S_z$, $\langle \hat{a}^\dagger \hat{a} \rangle$ and $\langle \hat{S}_+ \hat{S}_- \rangle$ (shown in Figs. 3a-3c) are consistent with results obtained from equation (2) when the masing condition (white curve in the figures) is satisfied. It is clearly seen that the photon number increases dramatically to a large value when the pump rate enters into the masing regime (Fig. 3b). Since the pump rate $w \sim O(N)$, the photon number scales with the number of spins by $n_c \approx \frac{w}{2\kappa_c}(N - S_z) \sim O(N^2)$, which demonstrates the superradiant nature of the maser. The fact that $\langle \hat{S}_+ \hat{S}_- \rangle \gg N_e$ unambiguously evidences phase correlation between a macroscopically large number of spins established by superradiant masing. The optimal pumping condition for correlation between spins is determined by maximizing the spin correlation

$$\langle \hat{S}_+ \hat{S}_- \rangle = S_z \left( \frac{w - \gamma_{eg}}{2\kappa_S} N - \frac{w + \gamma_{eg}}{2\kappa_S} S_z \right). \qquad (4)$$

Considering that under strong pump $w \gg 1/T_2^* \gg \gamma_{eg}$ and hence $\kappa_S \approx w$, the maximum collective spin correlation is reached when the pump rate is $w_{\text{opt}}^{\text{max-corr}} \approx 2Ng^2/\kappa_c$, where $S_z \approx N/2$, $\langle \hat{S}_+ \hat{S}_- \rangle \approx N^2/8$, and $n_c = \langle \hat{a}^\dagger \hat{a} \rangle = N^2 g^2/(2\kappa_c^2)$.

The coherence time of the macroscopic quantum superposition of magnons and photons is determined by the maser linewidth. We use the standard procedure to calculate the maser linewidth from the correlation of the phase fluctuations of photons or equivalently that of magnons, namely, $\langle [\delta\hat{a}(t_1) - \delta\hat{a}^\dagger(t_1)][\delta\hat{a}(t_2) - \delta\hat{a}^\dagger(t_2)]\rangle$ or $\langle [\delta\hat{S}_-(t_1) - \delta\hat{S}_+(t_1)][\delta\hat{S}_-(t_2) - \delta\hat{S}_+(t_2)]\rangle$. The coherence time is obtained as

$$T_{\text{coh}} = 4(\kappa_c^{-1} + \kappa_S^{-1})(n_c + n_S)/n_{\text{incoh}}, \tag{5}$$

where $n_c = \langle \hat{a}^\dagger \hat{a} \rangle$ is the photon number, $n_S = \langle \hat{S}_+ \hat{S}_- \rangle / S_z$ is the magnon number, and $n_{\text{incoh}} = n_{\text{th}} + N_e/S_z$ includes the thermal photon number ($n_{\text{th}}$) at temperature $T$ and the incoherent magnon number ($\langle \hat{S}_+ \hat{S}_- \rangle / S_z = \sum_{j=1}^N \langle \hat{s}_j^+ \hat{s}_j^- \rangle / S_z = N_e/S_z$ if the correlation between different spins are set to zero). Since each incoherent photon or magnon contributes a random phase $\sim\pi$, the coherence time is reduced by the incoherent population as compared with the intuitive result in equation (1). The coherence time is indeed greatly enhanced under the masing condition (Fig. 3d). For NV center spins in diamond, the magnon decay rate $\kappa_S > 10^6 \text{ sec}^{-1}$ while for a good microwave cavity ($Q > 10^5$) the photon decay rate $\kappa_c < 3\times 10^4 \text{ sec}^{-1}$. Thus the photon number $n_c = n_s \kappa_S/\kappa_c$ is much greater than the magnon number and the macroscopic quantum coherence is mainly maintained by the photons in the cavity. For a cavity with quality factor $Q = 10^5$ and a laser pump rate $w = 10^5 \text{ sec}^{-1}$ (marked by a green cross in Fig. 3), which are readily realizable, the coherence time is as long as $6.3 \times 10^7$ sec. The optimal pump condition for long coherence time can be obtained from equation (5). In the good-cavity or large ensemble limit where $\kappa_c/2g^2 \ll NT_2^*$, the optimal pump rate for maximum coherence time is approximately the same as for maximum spin-spin correlation, i.e., $w_{\text{opt}}^{\text{max-corr}} \approx 2Ng^2/\kappa_c$ and the optimal coherence time is

$$T_{\text{coh}}^{\text{opt}} \approx 4N^2 g^2 /(3\kappa_c^3). \tag{6}$$

The maximum coherence time of the macroscopic quantum state scales with the spin number and the cavity $Q$ factor by $T_{\text{coh}}^{\text{opt}} \propto N^2 Q^3$.

We also study the temperature dependence of the macroscopic quantum coherence. At higher temperature the coherence time is reduced due to increase of the spin relaxation rate [28] and the incoherent thermal photons. Since the spin relaxation rate $\gamma_{eg}$ is always much smaller than the other decay rates ($w$ and $2/T_2^*$), the dominating temperature effect is due to the thermal photons. The coherence times are shown in Fig. 4 for two higher temperatures, namely, 4 K and 300 K. At 300 K, the cavity has a large number of thermal photons ($n_{th} = 2083$), and the coherence time is reduced from the 120 mK value by ~3 orders of magnitude. Nonetheless, the macroscopic superposition still has a long coherence time ($7.4 \times 10^4$ sec for $Q = 10^5$ and $w = 10^5$ sec$^{-1}$, marked by a green cross in Fig. 4) even at room temperature. Such long coherence times are much longer than the $T_1$ time of the spins and would be limited only by the system stability (positions of the mirrors and the sample holders, the external magnetic field, the pumping rate, etc.).

The long coherence time of the macroscopic quantum superposition is a useful resource for quantum technologies, such as ultrasensitive magnetometry [29]. When the external magnetic field or the mirror position is changed such that the spin transition frequency is shifted away from the exact resonance with the cavity photon, the masing frequency will be dragged to $\omega = (\kappa_c \omega_S + \kappa_S \omega_c)/(\kappa_c + \kappa_S)$ [20]. The ultralong coherence time of the superradiant maser means ultranarrow linewidth and hence ultrasensitivity to the external magnetic field and the mirror position (see Fig. 3e & f). The sensitivity of a magnetic field with frequency $\leq (\kappa_c + \kappa_S)/2$ is estimated to be $\delta B \sqrt{t_m} = \gamma_{NV}^{-1}(1 + \kappa_S/\kappa_c)\sqrt{2T_{coh}^{-1}}$ for measurement time $t_m$, where $\gamma_{NV}/2\pi = 2.8 \, \text{MHz} \cdot \text{Gauss}^{-1}$ is the NV center gyromagnetic ratio. Thus the magnetic field sensitivity can reach up to $23 \, \text{fT} \cdot \text{Hz}^{-1/2}$ for $Q = 10^5$ and $w = 10^5$ sec$^{-1}$ at 120 mK, and $0.67 \, \text{pT} \cdot \text{Hz}^{-1/2}$ even at room temperature. The sensitivity to the cavity mirror position, $\delta x \sqrt{t_m} = (L/\omega_c)(1 + \kappa_c/\kappa_S)\sqrt{2T_{coh}^{-1}}$, is reduced by the large cavity length (L=50 mm). For the cavity quality factor $Q = 10^5$ and pump rate $w = 10^5$ sec$^{-1}$, the mirror position sensitivity $\delta x \sqrt{t_m} = 0.5 \, \text{fm} \cdot \text{Hz}^{-1/2}$ at 120 mK or $14 \, \text{fm} \cdot \text{Hz}^{-1/2}$ at room temperature. However, in the high-$Q$ regime, the position sensitivity is greatly enhanced while the magnetometry sensitivity is reduced due to the frequency dragging effect (see Fig. 3e & f). The sensitivities to the magnetic field and the mirror positions set the requirements on stability of the setup for maintaining the long coherence times of the macroscopic quantum superposition.

## METHODS SUMMARY

The theoretical study is based on the standard Langevin equations [22]

$$\begin{aligned}
\frac{d\hat{N}_e}{dt} &= +w\hat{N}_g - \gamma_{eg}\hat{N}_e + ig\left(\hat{a}^\dagger \hat{S}_- - \hat{S}_+ \hat{a}\right) + \hat{F}_e, \\
\frac{d\hat{N}_g}{dt} &= -w\hat{N}_g + \gamma_{eg}\hat{N}_e - ig\left(\hat{a}^\dagger \hat{S}_- - \hat{S}_+ \hat{a}\right) + \hat{F}_g, \\
\frac{d\hat{S}_-}{dt} &= -i\omega_S \hat{S}_- - \frac{\kappa_S}{2}\hat{S}_- + ig\left(\hat{N}_e - \hat{N}_g\right)\hat{a} + \hat{F}_S, \\
\frac{d\hat{a}}{dt} &= -i\omega_c \hat{a} - \frac{\kappa_c}{2}\hat{a} - ig\hat{S}_- + \hat{F}_c,
\end{aligned} \tag{6}$$

where $\hat{F}_{c/S/e/g}$ is the noise that causes the decay of the photons (c), the magnons (S), the population in the excited state (e), or that in the ground state (g). Note that the total spin number is written as an operator $\hat{N}$ to take into account the fluctuation due to population of the third spin state $|+1\rangle$. The population fluctuation, however, has no effect on the phase fluctuation of the maser.

By replacing the operators with their expectation values, we obtain the mean-field equations for the masing process at the steady state

$$\begin{aligned}
0 &= wN_g - \gamma_{eg}N_e + ig(a^* S_- - S_+ a), \\
0 &= i(\omega - \omega_S)S_- - \frac{\kappa_S}{2}S_- + igS_z a, \\
0 &= i(\omega - \omega_c)a - \frac{\kappa_c}{2}a - igS_-,
\end{aligned} \tag{7}$$

from which the masing frequency, the field amplitudes, and the spin polarization can be straightforwardly calculated.

The coherence time and linewidth are calculated using the spectrum of the phase fluctuations. The equations for the fluctuations are linearized, which is justified since the fluctuations are much smaller than the expectation values when masing occurs. The linearized equations are

$$\frac{d\delta\hat{N}_e}{dt} = +w\delta\hat{N}_g - \gamma_{eg}\delta\hat{N}_e + ig\left(S_-\delta\hat{a}^\dagger - S_+\delta\hat{a}\right) + ig\left(a^*\delta\hat{S}_- - a\delta\hat{S}_+\right) + \hat{F}_e,$$

$$\frac{d\delta\hat{N}_g}{dt} = -w\delta\hat{N}_g + \gamma_{eg}\delta\hat{N}_e - ig\left(S_-\delta\hat{a}^\dagger - S_+\delta\hat{a}\right) - ig\left(a^*\delta\hat{S}_- - a\delta\hat{S}_+\right) + \hat{F}_g,$$

$$\frac{d\delta\hat{S}_-}{dt} = -\frac{\kappa_S}{2}\delta\hat{S}_- + igS_z\delta\hat{a} + iga\left(\delta\hat{N}_e - \delta\hat{N}_g\right) + \hat{F}_S, \quad (8)$$

$$\frac{d\delta\hat{a}}{dt} = -\frac{\kappa_c}{2}\delta\hat{a} - ig\delta\hat{S}_- + \hat{F}_c.$$

By Fourier transform of these equations, the spectrum of the phase noise can be calculated and hence the maser linewidth is determined.

To investigate the correlations in both the masing and the spontaneous emission regimes, we derive the equations of motion for the correlation functions and take the expectation values of the relevant operators. That leads to

$$\frac{d\langle\hat{N}_e\rangle}{dt} = +w\langle\hat{N}_g\rangle - \gamma_{eg}\langle\hat{N}_e\rangle + ig\left(\langle\hat{a}^\dagger\hat{S}_-\rangle - \langle\hat{S}_+\hat{a}\rangle\right),$$

$$\frac{d\langle\hat{N}_g\rangle}{dt} = -w\langle\hat{N}_g\rangle + \gamma_{eg}\langle\hat{N}_e\rangle - ig\left(\langle\hat{a}^\dagger\hat{S}_-\rangle - \langle\hat{S}_+\hat{a}\rangle\right),$$

$$\frac{d\langle\hat{a}^\dagger\hat{S}_-\rangle}{dt} = -\frac{\kappa_S + \kappa_c}{2}\langle\hat{a}^\dagger\hat{S}_-\rangle + ig\left[\left(1-\frac{1}{N}\right)\langle\hat{S}_+\hat{S}_-\rangle + \langle\hat{N}_e\rangle + \langle\hat{a}^\dagger\hat{a}\rangle\langle\hat{S}_z\rangle\right], \quad (9)$$

$$\frac{d\langle\hat{S}_+\hat{S}_-\rangle}{dt} = -\kappa_S\langle\hat{S}_+\hat{S}_-\rangle - ig\langle\hat{S}_z\rangle\left(\langle\hat{a}^\dagger\hat{S}_-\rangle - \langle\hat{S}_+\hat{a}\rangle\right),$$

$$\frac{d\langle\hat{a}^\dagger\hat{a}\rangle}{dt} = -\kappa_c\langle\hat{a}^\dagger\hat{a}\rangle - ig\left(\langle\hat{a}^\dagger\hat{S}_-\rangle - \langle\hat{S}_+\hat{a}\rangle\right) + \kappa_c n_{th},$$

Here to make the equations close, we have used the approximation $\langle\hat{a}^\dagger\hat{a}\hat{S}_z\rangle \approx \langle\hat{a}^\dagger\hat{a}\rangle\langle\hat{S}_z\rangle$, $\langle\hat{a}^\dagger\hat{S}_z\hat{S}_-\rangle \approx \langle\hat{S}_z\rangle\langle\hat{a}^\dagger\hat{S}_-\rangle$, and $\langle\hat{S}_+\hat{S}_z\hat{a}\rangle \approx \langle\hat{S}_z\rangle\langle\hat{S}_+\hat{a}\rangle$, neglecting the higher-order correlations, which is well justified for Gaussian fluctuations.

**Acknowledgements** This work was supported by Hong Kong Research Grants Council Project No. 14303014 & The Chinese University of Hong Kong Focused Investments Scheme.

**Author Contributions** R.B.L. conceived the idea and supervised the project, R.B. L. & L.J. formulated the theory, R.B.L., L. J., S. Y. & J.W. designed the physical system, L.J. carried out the study, R.B.L. & J.L. wrote the paper, and all authors commented on the manuscript.

**Author Information** The authors declare no competing financial interests. Correspondence and requests for materials should be addressed to R.B.L. (rbliu@phy.cuhk.edu.hk).

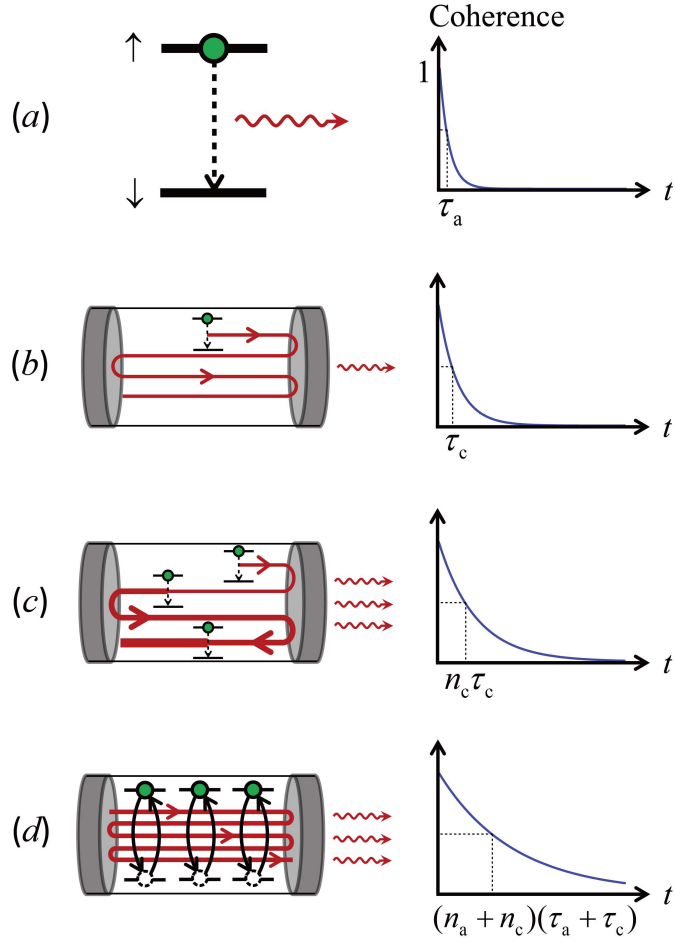

**Figure 1 | Schematic illustration of coherence time enhancement by superradiant lasing.** (a) In uncorrelated spontaneous emission of individual atoms, different photons have random phases, and hence the coherence time is determined by the atomic lifetime $\tau_a$. (b) The photon in a cavity after emission is reflected between mirrors and can stay in the cavity for a long time. The photons separated by a time longer than the cavity lifetime $\tau_c$ have phases uncorrelated. Therefore the photon coherence time is $\tau_c$. (c) In lasers, a large number ($n_c$) of photons are stored in the cavity due to stimulated emission, which have the same phase. The photon coherence time of the laser is therefore elongated to $n_c \tau_c$. (d) In superradiant lasing, the quantum coherence can be stored in both the cavity mode and the atomic collective mode, and the cavity photons and the collective mode excitations share the same phase. So the coherence time is enhanced to be $(n_a + n_c) \times (\tau_a + \tau_c)$, where $n_a$ denotes the number of atomic collective excitations.

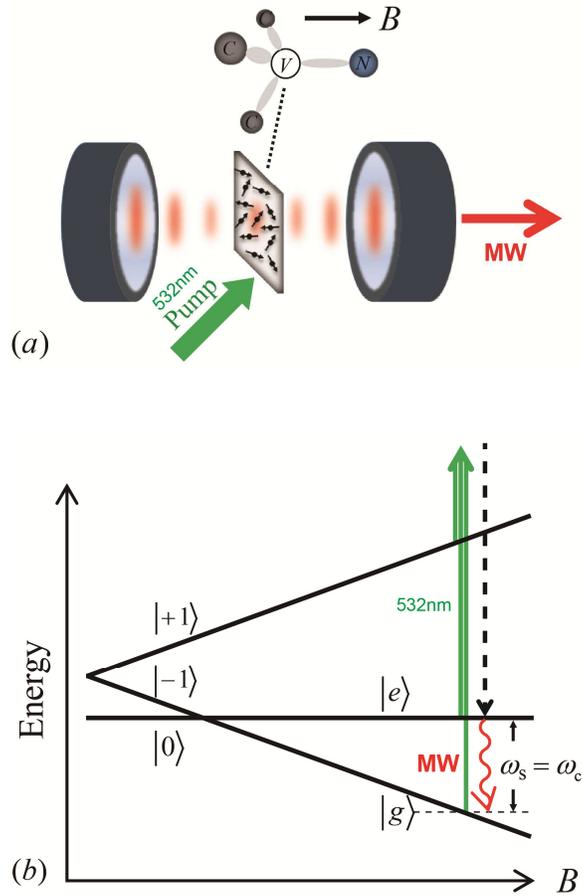

**Figure 2 | Schematic of superradiant masing in a spin ensemble coupled to a high-quality microwave cavity.** (a) System for superradiant masing. A diamond sample is fixed inside a high quality Febry-Pérot microwave cavity. A magnetic field is applied along the NV axis, which is set parallel to the cavity axis. The NV centers are pumped by a 532 nm laser (green arrow). (b) The energy levels of an NV center spin as functions of a magnetic field $B$. The zero-field splitting at $B=0$ is about 2.87 GHz. The magnetic field is set such that the transition frequency between the states $|-1\rangle$ ($|g\rangle$) and $|0\rangle$ ($|e\rangle$) is resonant with the cavity mode. The 532 nm light (green arrow) optically pumping the NV centers to the state $|e\rangle$, inducing a population inversion.

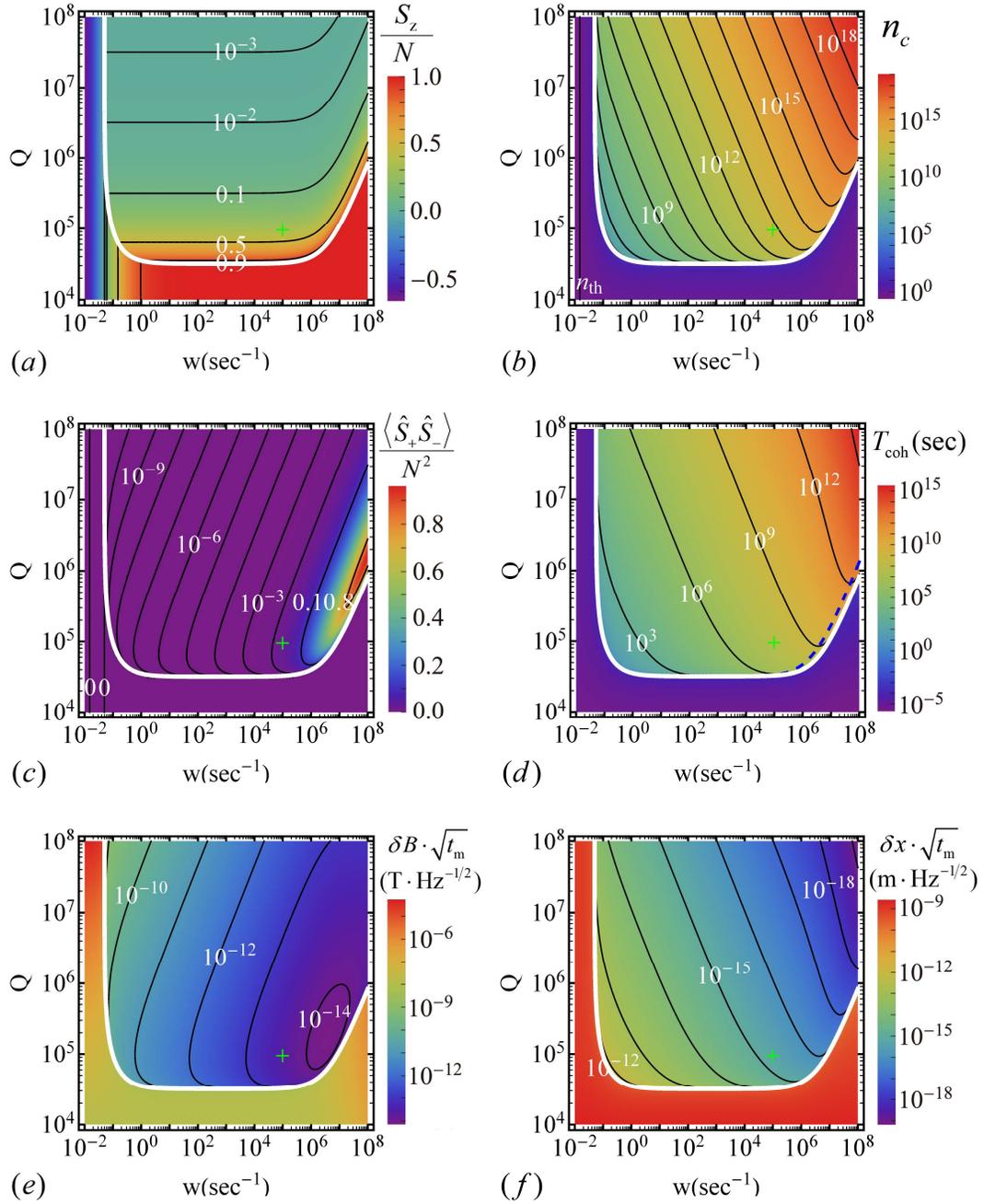

**Figure 3 | Macroscopic quantum coherence via superradiant masing.** (a) The spin polarization $S_z$, (b) the cavity photon number $n_c$, (c) the collective spin correlation $\langle \hat{S}_+ \hat{S}_- \rangle$, (d) the collective quantum coherence time $T_{coh}$, (e) sensitivity on the external magnetic field, and (f) sensitivity on the mirror position, as functions of the cavity $Q$ factor and the pump rate $w$. The masing condition for the pump rate $w$ is indicated in the figures by the white curves. The blue dashed curve in (d) shows the optimal pump

condition for maximum coherence time. The green crosses in the figures mark the point for $Q = 10^5$ and $w = 10^5 \text{sec}^{-1}$. The parameters are such that $\omega_c/2\pi = \omega_s/2\pi = 3$ GHz, $T_2^* = 0.5$ μs, $N = 0.375 \times 10^{14}$, at the temperature is 120 mK, the effective coupling is $g/2\pi = 0.02$ Hz, and the spin relaxation rate is $\gamma_{eg} = 0.05 \text{ sec}^{-1}$.

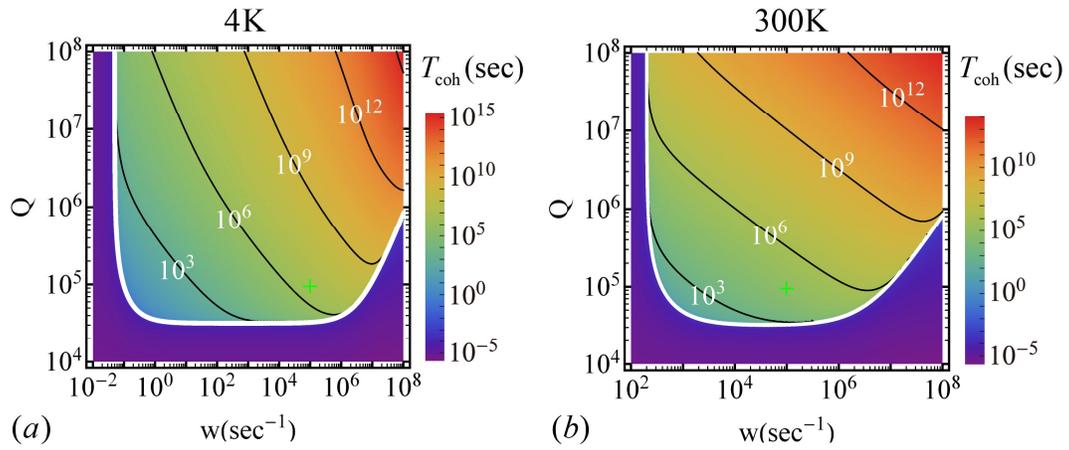

**Figure 4 | Temperature dependence of collective quantum coherence time.** The coherence time $T_{coh}$ is plotted as functions of the cavity $Q$ factor and the pump rate $w$ for temperature equal to (a) 4 K and (b) 300 K. The green crosses in the figures mark the point for $Q = 10^5$ and $w = 10^5 \text{ sec}^{-1}$. The spin relaxation rate $\gamma_{eg} = 0.05 \text{ sec}^{-1}$ at 4 K and $200 \text{ sec}^{-1}$ at 300 K. The other parameters are the same as in Fig. 3.